**Giant Planet Observations with the James Webb Space Telescope**

James Norwood

Dept. of Astronomy

Box 30001/MSC 4500

New Mexico State University

Las Cruces, NM 88003-0001, USA

jnorwood@nmsu.edu

Julianne Moses

Space Science Institute

4750 Walnut St., Suite 205

Boulder, CO 80301, USA

jmoses@spacescience.org

Leigh N. Fletcher

Department of Physics and Astronomy

University of Leicester

University Road

Leicester, LE1 7RH, United Kingdom

leigh.fletcher@le.ac.uk

Glenn Orton

MS 183-501

Jet Propulsion Laboratory

California Institute of Technology

4800 Oak Grove Drive

Pasadena, California 91109, USA

Glenn.S.Orton@jpl.nasa.gov

Patrick G. J. Irwin

Atmospheric, Oceanic, and Planetary Physics

Department of Physics

University of Oxford

Clarendon Laboratory

Parks Road, Oxford OX1 3PU, United Kingdom

patrick.irwin@physics.ox.ac.uk

Sushil Atreya

Department of Atmospheric, Oceanic, and Space Sciences

University of Michigan

Ann Arbor, MI 48109, USA

atreya@umich.edu

Kathy Rages

SETI Institute

189 Bernardo Ave.

Mountain View, CA 94043, USA

krages@seti.org

Thibault Cavalié

LESIA

Observatoire de Paris, CNRS, UPMC

Université Paris-Diderot

5 place Jules Janssen

92195 Meudon, France

thibault.cavalie@obspm.fr

Agustin Sánchez-Lavega

Física Aplicada I

Universidad del País Vasco UPV/EHU

ETS Ingeniería

Alameda Urquijo s/n, 48013, Bilbao, Spain

ricardo.hueso@ehu.es

Ricardo Hueso

Física Aplicada I

Universidad del País Vasco UPV/EHU

ETS Ingeniería

Alameda Urquijo s/n, 48013, Bilbao, Spain

ricardo.hueso@ehu.es

Nancy Chanover

Dept. of Astronomy

Box 30001/MSC 4500

New Mexico State University

Las Cruces, NM 88003-0001, USA

nchanove@nmsu.edu

## Abstract

This white paper examines the benefit of the upcoming James Webb Space Telescope for studies of the Solar System's four giant planets: Jupiter, Saturn, Uranus, and Neptune. JWST's superior sensitivity, combined with high spatial and spectral resolution, will enable near- and mid-infrared imaging and spectroscopy of these objects with unprecedented quality. In this paper we discuss some of the myriad scientific investigations possible with JWST regarding the giant planets. This discussion is preceded by the specifics of JWST instrumentation most relevant to giant planet observations. We conclude with identification of desired pre-launch testing and operational aspects of JWST that would greatly benefit future studies of the giant planets.

## Introduction

The upcoming James Webb Space Telescope will provide a tremendous advancement in near- and mid-infrared astronomy, with sensitivity and spectral resolution greatly superior to any of its predecessors. Observations with JWST will greatly benefit scientific investigations in numerous fields of astronomy, and studies of the Solar System's four giant planets are no exception, with the increased sensitivity and spatial resolution improving our understanding of Uranus and Neptune in particular. In this white paper, we examine JWST observations of the giant planets in detail, focusing on methodology and key science questions answerable by JWST. We also identify several unique observing requirements that might not necessarily apply to other JWST observation campaigns.

## Observing the Giant Planets

Unlike most of the targets that motivated JWST, the giant planets are bright, extended objects. This acts as a double-edged sword: these planets are bright enough for each instrument's highest spectral-resolution mode to be ideal (if not mandatory) for observing, but they run the risk of surpassing JWST's saturation limits. To ameliorate this situation, sub-array readout patterns are being implemented on JWST instruments, to reduce the minimum integration time and thus raise the saturation limits. With such methods in place, Uranus and Neptune will be observable throughout JWST's entire spectral range (0.7-28.5 μm); Jupiter and Saturn are expected to be observable throughout the near infrared, but will still saturate in parts of the mid-infrared. The brightnesses of all four planets compared to these higher saturation limits are presented in Figures 1-3.

The optimal observing configurations for the giant planets are detailed below.

- Near-infrared imaging with NIRCam: The pixel scales of NIRCam are 0.032″ (0.6-2.5 μm) and 0.065″ (2.5-5.0 μm), generally smaller than the diffraction limit (0.025″ at 0.6 μm, 0.105″ at 2.5 μm, and 0.21″ at 5 μm). Sub-array imaging will be required for some filters. The 640×640 sub-array shown in Figure 4 will provide efficient coverage of Jupiter, and will reduce the minimum exposure time enough for Jupiter to be observable in most medium filters. The smaller 400×400 sub-array is similarly suited for Saturn's disk and will make all medium filters useable for that planet. The smaller sub-arrays available will make both planets observable in all NIRCam filters, though with greatly reduced coverage. Uranus and Neptune are dim enough not to require sub-arrays for almost all filters.

- Near-infrared spectroscopy with NIRSpec: For most giant planet observations, NIRSpec is best used in Integrated Field Unit (IFU) mode at maximum spectral resolution ($R$ = 2700). This will provide spectra from a 3″×3″ field of view with a sampling of 0.1″ (coarser than the diffraction limit shortward of 2.5 μm, finer than the diffraction limit longward of 2.5 μm). The IFU's field of view is a little smaller than the angular size of the Jupiter's Great Red Spot, and mosaicking will be necessary to view the entire vortex (let alone the entire Jovian disk, roughly 40″ in diameter). Neptune is the only one of the four giant planets for which full-disk spectroscopy will be possible without mosaicking. As shown in Figure 2, Jupiter will likely be too bright shortward of 1.7 μm for the NIRSpec IFU, though it will be well below the saturation limit of NIRSpec's 0.2″ × 3.3″ slit (also with a maximum spectral resolution of $R$ = 2700) for most of this spectral region.

- Mid-infrared imaging with MIRI: The pixel scale of MIRI's imager is 0.11″, well below the diffraction limit. Its largest sub-arrays have a field of view of 7″, approximately twice the angular diameter of Uranus. As shown in Figure 3, Uranus and Neptune will be observable in all nine MIRI filters. Saturn may be imaged in filters shortward of 12 μm. Its summer hemisphere will saturate the F1280W and F1500W filters, but closer to equinox imaging may be possible out to 16 μm. (Saturn's winter hemisphere would be obscured by the rings.) Unfortunately, mid-infrared imagery of Jupiter will be almost impossible: only Jovian zones are expected to be dark enough, and only for MIRI's F560W filter (Norwood et al. 2015, Figure 7).

- Mid-infrared spectroscopy with MIRI's Medium-Resolution Spectrometer (MRS): MRS has four IFUs. The shortest-wavelength IFU (5.0-7.71 μm) has a 3.0″ × 3.9″ field of view (comparable in angular size to Uranus, as shown in Fig. 9 of Norwood et al. 2015), a spectral resolution of $R$=3250, and sampling of 0.19″. The longest-wavelength IFU (18.35-28.3 μm) has a 6.7″ × 7.7″ field of view, with $R$=1550 and 0.28″ sampling. The spatial resolutions of all four IFUs, though better than the diffraction limit, do not offer Nyquist sampling of the diffraction limit. Thus, MRS offers greater spectral resolution than the imager in exchange for slightly less spatial information. Uranus and Neptune will be observable over the full mid-IR spectrum, while Jupiter and Saturn will saturate longward of ~10 and ~20 μm, respectively.

JWST's field of regard covers solar elongation angles from 85° to 135°, resulting in each planet having a ~50-day observing window occurring approximately every six months. Of these four planets, Jupiter has the largest rate of movement at a maximum of 4.5 mas/second, well within JWST's maximum tracking rate of 30 mas/second. For moving targets, JWST's pointing stability specification is 0.050″ (3σ). JWST's roll range is +/-5° about the telescope's optical axis, granting little freedom in the orientation of the imager/spectrometer relative to that of the planet.

For observations devoted to individual features rather than the entire planetary disk, the target coordinates will be calculated in terms of offset relative to the planet's ephemeris. A feature's position will be extrapolated from the most recent imaging available, using the planet's rotation rate, zonal wind speeds, and the most recent observed drift rate of the feature. (For Jupiter and Saturn, the amateur community can often produce recent images for this purpose.) If there is uncertainty in the target's position, multiple observations may be implemented, scanning the approximate neighborhood. The feature's location at the time of the observations may be confirmed using images either immediately before or immediately after the primary observations.

**Uranus and Neptune**

Perhaps the greatest contribution that JWST will make to giant-planet studies will be spatially resolved observations of the ice giants Uranus and Neptune in the mid-infrared. The mid-infrared spectra of these planets are replete with stratospheric emission features, primarily from photochemically produced hydrocarbons such as ethane (12.2 μm) and acetylene (13.7 μm). Several stratospheric hydrocarbons have been detected on these planets, including $C_2H_6$, $C_2H_2$, $C_3H_4$, $C_4H_2$, and (on Neptune only) $C_2H_4$ and $CH_3$ . The latest assessment of stratospheric hydrocarbons came from full-disk Spitzer observations (Orton et al. 2014); the superior sensitivity and spectral resolution of JWST will greatly improve this inventory, and facilitate detection of additional hydrocarbons for which only upper limits are known (e.g. $CH_3$ and $C_2H_4$ in Uranus). JWST will be particularly valuable for investigating species that cannot be observed from the ground due to telluric absorption, such as $C_3H_4$, $C_4H_2$, $C_6H_6$, and $CH_3$. Some species, such as $C_3H_8$ and $C_2H_4$, will still be difficult to observe due to their signatures being hidden within emission from other hydrocarbons, though they may be identified in the residuals when fitting the stronger features. By sounding stratospheric temperature from methane and hydrogen emission at the same time as composition, JWST will help eliminate the temperature-abundance degeneracies in analysis of these planets' mid-infrared spectra, and will refine existing models of the photochemistry, vertical motion, and circulation occurring in these planets' stratospheres (e.g., Moses et al. 2005, Dobrijevic et al. 2010). Repeated observations will also clarify the nature of previously observed variability in this emission (Hammel et al. 2006, Fletcher et al. 2014), such as whether these changes are driven primarily by seasonal variations, solar activity, or short-timescale weather patterns.

Particularly important is that spatial distributions of these species have not yet been adequately investigated: Greathouse et al. (2011) and Fletcher et al. (2014) present meridional distributions of $C_2H_6$ and $C_2H_2$ on Neptune, but the spatial resolution of MIRI will permit a much more comprehensive mapping of emission from a broader range of stratospheric hydrocarbons. These maps will provide insights into the circulation and thermal structure within the stratospheres and upper tropospheres of these planets, and the coupling between the convective weather layers and the stable stratosphere. By characterizing the spatial scales involved, these maps will clarify the role weather systems play (if any) in stratospheric thermal structure and abundances. Lastly, spectroscopy with high spatial resolution will enable center-to-limb comparisons, which will grant additional vertical information for temperature and abundance profiles.

Also observable in these planets' spectra are stratospheric species that may be external in origin: emission from CO2, CO, and HCN. 15-μm emission from $CO_2$ was previously detected on Uranus using

Spitzer/IRS (Orton et al. 2014), and 4.7-μm fluorescent emission from CO was observed on Neptune from AKARI (Fletcher et al. 2010). JWST will be able to observe these features at higher spectral resolution and much greater SNR. Characterization of the abundance, temperature and spatial distribution of this emission will place tighter constraints upon the sources of stratospheric oxygen: infalling ring particles, material from satellites, impacts from comets or Kuiper Belt objects (Lellouch et al. 2005, Cavalié et al. 2014), or upwelling from below. Similarly, HCN was detected in the aftermath of the Shoemaker-Levy 9 impacts on Jupiter (e.g., Lellouch et al. 2006), and its detection on Neptune (Rosenqvist et al. 1992, Marten et al. 1993) also suggests an external origin (Marten et al. 2005, Rezac et al. 2014). If JWST is able to map HCN on Neptune, it would clarify whether the origin was a single event (such as a large impact) or gradual (e.g., dust from the Kuiper belt). JWST will also be able to determine upper limits for stratospheric HCN on Uranus.

In the near infrared, one of the key studies will be investigation of the composition and vertical structure of these planets' upper tropospheres. In this wavelength regime, the sensing depth for Uranus and Neptune is dependent primarily on the opacity of methane, which varies over several orders of magnitude and thus allows observations to probe to a variety of altitudes. Multi-wavelength and center-to-limb observations will be able to locate and characterize the cloud layers from which incident sunlight is reflected: the uppermost ($CH_4$) cloud layer, and if it is optically thin enough the (presumably $H_2S$) cloud below it. However, a new complexity in such a method was revealed with the discovery of latitudinal variations in the methane abundance on both planets (Karkoschka and Tomasko 2009, 2011; Sromovsky et al. 2011, 2014). JWST observations will be able to carefully quantify the methane abundance as a function of latitude by comparing the strength of methane absorption in these planets' near-IR spectra to collision-induced hydrogen absorption, which becomes prominent near 0.825 μm. Such an investigation will not only benefit future modeling reliant on $CH_4$ opacity, but may also clarify the depth to which this variation persists and its implications for global circulation models. Investigations of the $CH_4$ in the stratosphere will also reveal whether its distribution varies with latitude as well, and will help assess the very different disk-averaged $CH_4$ profiles derived from Spitzer and Herschel observations (Lellouch et al. 2015).

Given the very long orbital periods of Uranus (84 years) and Neptune (165 years), studies of these planets' seasonal variations are a challenging endeavor, and while full seasonal cycles are outside the mission lifespan of JWST, it will be possible to capture important views of states that will not return again for decades, contributing to the robust baseline needed to assess seasonal behavior and distinguish it from shorter-term weather phenomena. As Uranus has been approaching solstice (2030, northern summer) and Neptune equinox (2046, northern spring), several signs of seasonal evolution have been seen even on five-year timescales, such as the appearance of a bright polar collar in Uranus' northern hemisphere as the southern counterpart has disappeared (Irwin et al. 2009, Sromovsky et al. 2015), and Neptune's south polar vortex changing in strength (Fletcher et al. 2014).

The scientific opportunities at Uranus and Neptune are numerous. Aside from the above explorations, other near- and mid-infrared investigations involving these planets include

- tracking of cloud features to determine zonal wind speeds. Such an investigation is best implemented with near-infrared imagery over timescales of 2-3 times the planet's rotational period (resulting in spans of 30-50 hours). A recent reexamination of Voyager 2 Uranus imagery by Karkoschka (2015) has found southern hemisphere wind profiles different from those currently seen in the northern hemisphere, pointing to hemispheric asymmetry and/or seasonal evolution.
- searching for the near-infrared signatures of as yet undetected tropospheric species. Prominent absorption features from $AsH_3$ and $GeH_4$ have been observed in the spectra of Jupiter and Saturn (e.g., Noll and Larson 1991), and the sensitivity of NIRSpec may allow their observation on Uranus and Neptune. Near-infrared features from $PH_3$ and HCN may be identified as well, but such detections will be difficult.

- near-infrared mapping of $H_3^+$ and $H_2$ quadrupole emission to study the thermospheres, aurorae, and magnetic fields of Uranus and Neptune.  Of particular interest is whether these oddly oriented magnetic fields have changed since the Voyager flybys.

- comparing $CH_3D$ absorption features to those of $CH_4$ to determine the D/H ratios of the ice giants' atmospheres, which will constrain the properties of the protoplanetary ices from which they mainly formed.  Of note, Feuchtgruber et al. (2013) found D/H ratios in these atmospheres lower than previous determinations; JWST measurements will assist in more accurate assessment.

- determination of the hydrogen ortho/para ratio as a function of latitude and altitude, which will improve our understanding of the conversion process between the two, and its role in energy balance and global circulation.  This quantity has not been measured since the Voyager era.

**Jupiter and Saturn**

Unlike the two ice giants, Jupiter and Saturn are too bright to be observed by JWST in much of the mid-infrared even with sub-array imaging.  Jupiter is mainly observable only with spectroscopy shortward of 10-11 μm (depending on local brightness).  Because of this, the hydrogen-helium collision-induced continuum is largely above saturation limits, as is stratospheric emission from acetylene (13.7 μm) and ethane (12.3 μm).  As a result, JWST studies of Jupiter will be limited primarily to reflected sunlight (<5 μm), and tropospheric composition in the 5-10 μm region.  The only observable hydrocarbon on Jupiter would be $CH_4$: its 7.7-μm emission will be useful in determining stratospheric temperatures, and its 3.3-μm emission will probe the $CH_4$ abundance near the homopause.  While most of these spectral regions are accessible from the ground, there are still several localities in the near infrared where telluric $H_2O$ and $CO_2$ absorption make space-based observations the only option.  Additionally, the excellent spectral resolution of JWST will provide higher-quality data than most ground-based facilities.

For Saturn, the saturation limits will allow spectroscopy over more of the mid-infrared, granting access to many more stratospheric features including $C_2H_6$ at 12.3 µm, HCN at 14 µm, $CO_2$ emission at 15 µm, and most of the hydrocarbons previously discussed for Uranus and Neptune. The 13.7-µm emission from the Q-branch of $C_2H_2$ will exceed saturation limits, but the other $C_2H_2$ emission features near it should still be observable. Other species of particular interest are CO, $C_3H_4$, $C_4H_2$, and $C_6H_6$, all with features unobservable from the ground. These species lie near the saturation cutoff, and their observability may be seasonally dependent: likely observable near equinox but not near solstice (when the summer hemisphere is hot and the winter hemisphere is obscured by rings). If observable with JWST, analysis of their features would be very useful for constraining photochemistry and circulation. Also beneficial will be observations of the aforementioned 7.7-µm and 3.3-µm methane emission. Temperature maps of the middle atmosphere of Saturn, for example, probe the most seasonally variable region of the atmosphere (Moses and Greathouse 2005, Fletcher et al. 2010b, Hue et al. 2015), and the location of large-scale wave phenomena in both the horizontal and vertical (Fouchet et al. 2008, Guerlet et al. 2011). These observations will help constrain the emerging general circulation models of these planets (Friedson and Moses 2012, Medvedev et al. 2013, Guerlet et al. 2014). Even though other investigations have already been mapping the more prominent emission features on these planets, JWST-produced maps will still have some advantages: better determination of absolute abundances due to radiometric calibration issues experienced by ground facilities, and observing at Saturnian seasons inaccessible to Cassini.

Numerous previously-detected tropospheric constituents such as $NH_3$, $PH_3$, $H_2O$, $CH_3D$, and $^{15}NH_3$ have observable signatures shortward of 12 µm, so the high spatial resolution and unsurpassed sensitivity of JWST observations will greatly improve upon current determinations of the vertical and horizontal distributions of these species in the atmospheres of Jupiter and Saturn. JWST may also have the spectral resolution to detect and map tropospheric $AsH_3$ and $GeH_4$. An improved assessment of tropospheric abundances and how they vary will greatly assist in our understanding of global and regional temperature, chemistry, and dynamics in these planets' atmospheres. The spectral region around the 5-µm window will be particularly valuable, with signatures of disequilibrium species like CO, $PH_3$, $AsH_3$, and $GeH_4$; as well as tropospheric $H_2O$ above the cloud (humidity measurements mainly of value to vertical-motion studies). Such investigations can also explore the contrast between the two hemispheres of Saturn, as well as characterize any evolution since the Cassini era (occurring at Saturn's 2017 solstice).

The larger angular sizes of Jupiter and Saturn will allow near-IR studies of atmospheric dynamics on these planets to far greater detail than on Uranus and Neptune. (For example, the NIRCam pixel scale would be 110, 220, 450, and 690 km at the sub-observer point for Jupiter, Saturn, Uranus, and Neptune, respectively.) Reflectivity observations that make use of the variable opacity of methane and center-to-limb variations will enable altitude determinations of the cloud features in storms and vortices, 1-3 μm mapping of $NH_3$-ice and $H_2O$-ice distributions (Baines et al. 2009, Sromovsky et al. 2013) will clarify their origins and dynamics, and repeat observations will explore the motions and evolution of these phenomena. Among the phenomena of interest are variable Jupiter's South Equatorial Belt and North Temperate Belt Disturbances (Sanchez-Lavega et al. 2008), Saturn's recurrent Great White Spots (Sanchez-Lavega et al. 2011, Sayanagi et al. 2013), and the ongoing evolution of Jupiter's Great Red Spot as it diminishes in size. JWST observations will also assist investigations of atmospheric evolution on these dynamic worlds as the events unfold.

Further atmospheric studies include observations of impact events such as the Shoemaker-Levy 9 encounter (if they occur when the planet is favorably positioned); JWST observations would be able to track the aftermath to investigate atmospheric composition, vertical structure, and dynamics. Near-infrared auroral studies on Jupiter and Saturn are also possible, including examination of auroral morphology, variability, and $H_3^+$ variability as a function of latitude and local time. Such observations would not be unique at Jupiter: Juno and JUICE will be performing such investigations, and they are possible (with difficulty) from the ground. At Saturn, studies of the influence of the rings upon the $H_3^+$ distribution will be possible.

Lastly, JWST observations of Saturn will greatly compliment the current rich dataset of the Cassini mission. While JWST will not match the spatial resolution afforded by Cassini, NIRSpec will offer superior spectral resolution ($R$ = 2700) compared to VIMS ($R$ ~ 200), and MIRI will fill in the gap in coverage between VIMS and CIRS (5-7 μm), giving access to additional absorption and emission features of atmospheric $CH_4$, $NH_3$ and $C_2H_6$. Furthermore, by the end of its mission Cassini will have explored the Saturnian system over half of a Saturnian year, roughly from southern summer solstice to northern summer solstice; JWST's anticipated lifespan will cover most of the second half of this Saturnian year, the two providing together a more complete look at a full seasonal cycle.

**Desired capabilities for giant-planet studies**

The most significant issue regarding giant-planet observations with JWST is that the targets are bright objects, and thus the saturation limits of JWST instruments will be more important to planning observations than their sensitivity thresholds. Precise determination of these saturation limits will be vital for planning observations of these objects, and any novel techniques that could raise these limits would greatly benefit future giant planet investigations. Since sub-array imaging will be essential for many desired giant planet observations, a thorough investigation of this observing process is needed to ascertain what steps, limitations, and drawbacks this method entails.

Because giant-planet observations will often approach the saturation limits, characterization of JWST's performance under "bright, extended object" conditions will be important. A confident recipe for such observations will likely be needed, plotting out which "cool-down" measures - such as nodding to empty sky for a set period of time and repositioning the target elsewhere within the field of view - would be needed between one bright observation and the next. (Subsequent observers would also appreciate not having residual effects from previous bright target observations.)  Planning observations will also require precise knowledge of the saturation limits of JWST's instruments.  Characterization of scattered light effects would determine the feasibility of observing dimmer regions on these planets that happen to lie near bright, possibly saturated regions.

As the giant planets are extended objects, multiple small-offset observations will play a significant role in observing strategies.  Since Jupiter and Saturn (and in some cases, Uranus) are greater than the field of view, a full planet view would require multiple images mosaicked together.  (Should mosaics turn out to be infeasible, smaller-scale comparisons would still be possible, such as belt-zone comparisons on Jupiter.)  Furthermore, for observing modes with pixel scales larger than the diffraction limit, observing strategies will achieve higher spatial resolution through dithering (sub-pixel offsets).  With these plans in mind, an observer would need JWST to be able to slew short distances repeatedly without the overhead entailed in larger slews.  In particular, the following items would be of interest to such observers:

- the ability to conduct short-distance slews for mosaics and bright-target nod-offs, without the full wait duration associated with long-distance slews.
- characterization of the time it takes the telescope to fully stabilize after slewing a short distance.
- pointing accuracy of ~8 milliarcseconds (about ~1/4 of the pixel scale of NIRCam, the instrument with the finest spatial resolution) or better while tracking a moving object, to enable dithering with half-pixel offsets.  In addition to properly placing the planet in the field of view, this knowledge will also assist in determining locations of features on the planet itself when the limb is not present.
- clarification of any additional measures necessary when taking numerous images of bright objects, as noted earlier.

Lastly, it must be noted that the giant planets have been known to exhibit events or changes over very short timescales.  When an impact occurs, observers are treated to unique views of atmospheric dynamics and chemistry, as the tracer molecules are circulated throughout the atmosphere and ultimately dissipate (e.g., Hueso et al. 2010, Orton et al. 2011). Studies of rapid changes in the belt-zone systems (such as the 2010 disappearance of Jupiter's South Equatorial Belt) and of short-lived storm systems (such as Saturn's giant 2010-2011 storm) in both the near- and mid-infrared can clarify the roles of the structural and thermal properties involved in these atmospheric processes. If such time-critical events are to be studied with JWST (when in the field of regard), it will be necessary to implement target-of-opportunity protocols that may be triggered with as little as tens of hours warning. Studies of impact aftermath are best conducted within 24 hours of discovery; storm development typically has a window of roughly one week, while vortex mergers can be predicted 1-2 months in advance.

REFERENCES

Baines et al., (2009), *Plan. Space Sci.*, 57, 1650-1658.

Burgdorf et al. (2008). BAAS 40, 489.

Cavalié et al. (2014). *A&A* 562, A33.

Clark and McCord (1979). *Icarus* 40, 180-188.

Dobrijevic et al. (2010). *P&SS* 58, 12, 1555-1566.

Encrenaz et al. (1997). Proc. First ISO Workshop on Analytical Spectroscopy. ESA-SP, 419 (1997), 125-130.

Feuchtgruber at al. (2013). *A&A* 551, A126.

Fink and Larson (1979). *ApJ* 1, 233, 3, 1021-1040.

Fletcher et al. (2014). *Icarus* 231, 146-167.

Fletcher et al. (2010a). *A&A* 514, A17.

Fletcher et al. (2010b). *Icarus* 208, 1, 337-352.

Fletcher et al. (2009). *Icarus* 202, 2, 543-564.

Fouchet et al. (2008). *Nature* 453, 7192, 200-202.

Friedson and Moses (2012). *Icarus* 218, 2, 861-875.

Greathouse et al. (2011). *Icarus* 214, 2, 606-621.

Guerlet et al. (2014). *Icarus* 238, 110-124.

Guerlet et al. (2011). *GRL* 38, 9, L09201.

Hammel et al. (2006). *ApJ* 644, 2, 1326-1333.

Hue et al. (2015). *Icarus* 257, 163-184.

Hueso et al. (2010). *ApJ* 721, 2, L129-L133.

Irwin et al. (2009). *Icarus* 203, 1, 287-302.

Karkoschka (2015). *Icarus* 250, 294-307.

Karkoschka (1994). *Icarus* 111, 1, 174-192.

Karkoschka and Tomasko (2011). *Icarus* 211, 1, 780-797.

Karkoschka and Tomasko (2009). *Icarus* 202, 1, 287-309.

Lellouch et al. (2015). *A&A* 579, A121.

Lellouch et al. (2006). *Icarus* 184, 2, 478-497.

Lellouch et al. (2005). *A&A* 430, L37-L40.

Marten et al. (2005). *A&A* 429, 1097-1105.

Marten et al. (1993). *ApJ* 406, 1, 285-297.

Medvedev et al. (2013). *Icarus* 225, 1, 228-235.

Moses et al. (2005). *JGR* 110, E08001.

Moses and Greathouse (2005). *JGR* 110, E09007.

Noll and Larson (1991). *Icarus* 89, 168-189.

Norwood et al. (2015). *PASP*, in revision.

Orton et al. (2014). *Icarus* 243, 494-513.

Orton et al. (2011). *Icarus* 211, 1, 587-602.

Rezac et al. (2014). *A&A* 563, A4.

Rosenqvist et al. (1992). *ApJ* 392, 2, L99-L102.

Sanchez-Lavega et al. (2008). *Nature,* 451, 437-440.

Sanchez-Lavega et al. (2011). *Nature* 475, 71-74.

Sayanagi et al. (2013). *Icarus* 223, 460-478.

Sromovsky et al. (2015). *Icarus* 258, 192-223.

Sromovsky et al. (2014). *Icarus* 238, 137-155.

Sromovsky et al. (2013). *Icarus* 226, 402-418.

Sromovsky et al. (2011). *Icarus* 215, 1, 292-312.

FIGURES

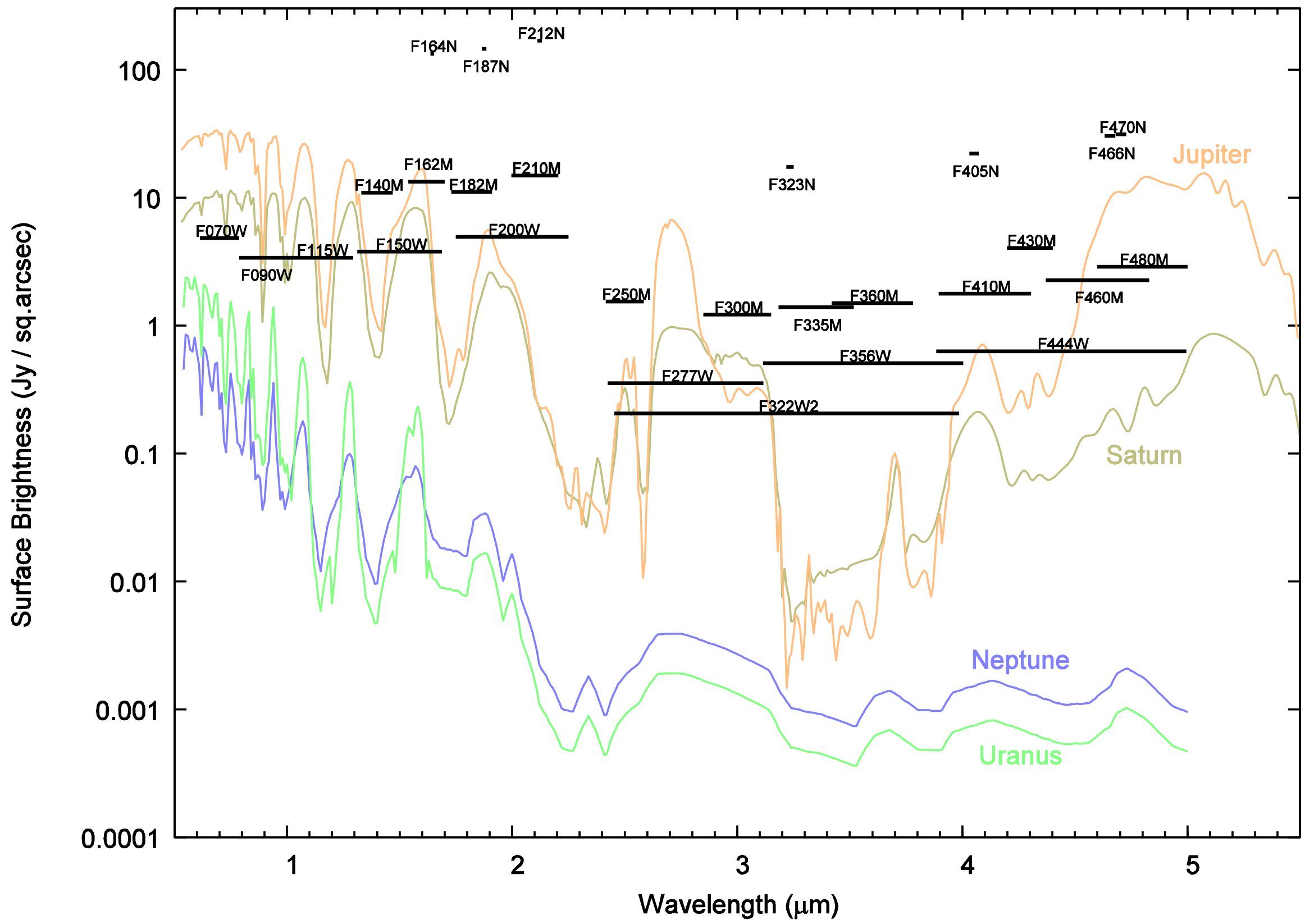

Figure 1. Near-infrared spectra of the four giant planets compared to the saturation limits of NIRCam filters, assuming 640×640 pixel sub-array imaging. As shown in Figure 4, the 640×640 sub-array is large enough to efficiently observe the full disk of Jupiter. Using the Saturn-sized 400×400 sub-array will raise the plotted saturation limits by a factor of ~2.5. The smallest sub-array (64×64) will raise the plotted limits by a factor of ~85, enough for Jupiter to be observable in all filters. Spectra composited from Clark and McCord (1979), Karkoschka (1994), Encrenaz et al. (1997), Fink and Larson (1979), and Burgdorf et al. (2008).

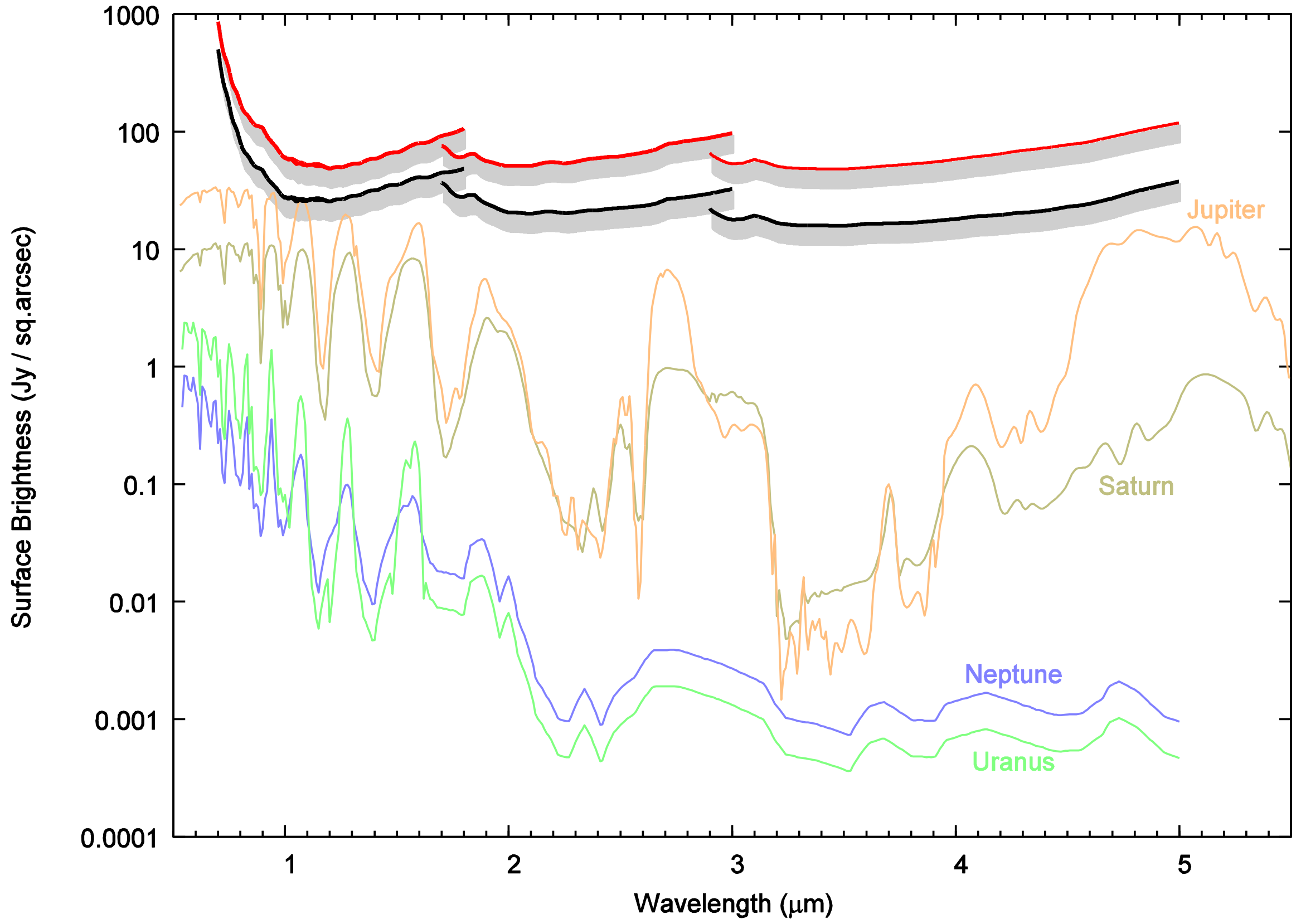

Figure 2. Near-infrared spectra of the four giant planets (same as in Figure 1) compared to saturation limits of the NIRSpec IFU (black) and slit (red). In both cases NIRSpec is in maximum spectral resolution mode ($R$=2700, several times greater than that of any of the spectra plotted here). The gray regions represent a 30% uncertainty in the saturation values.

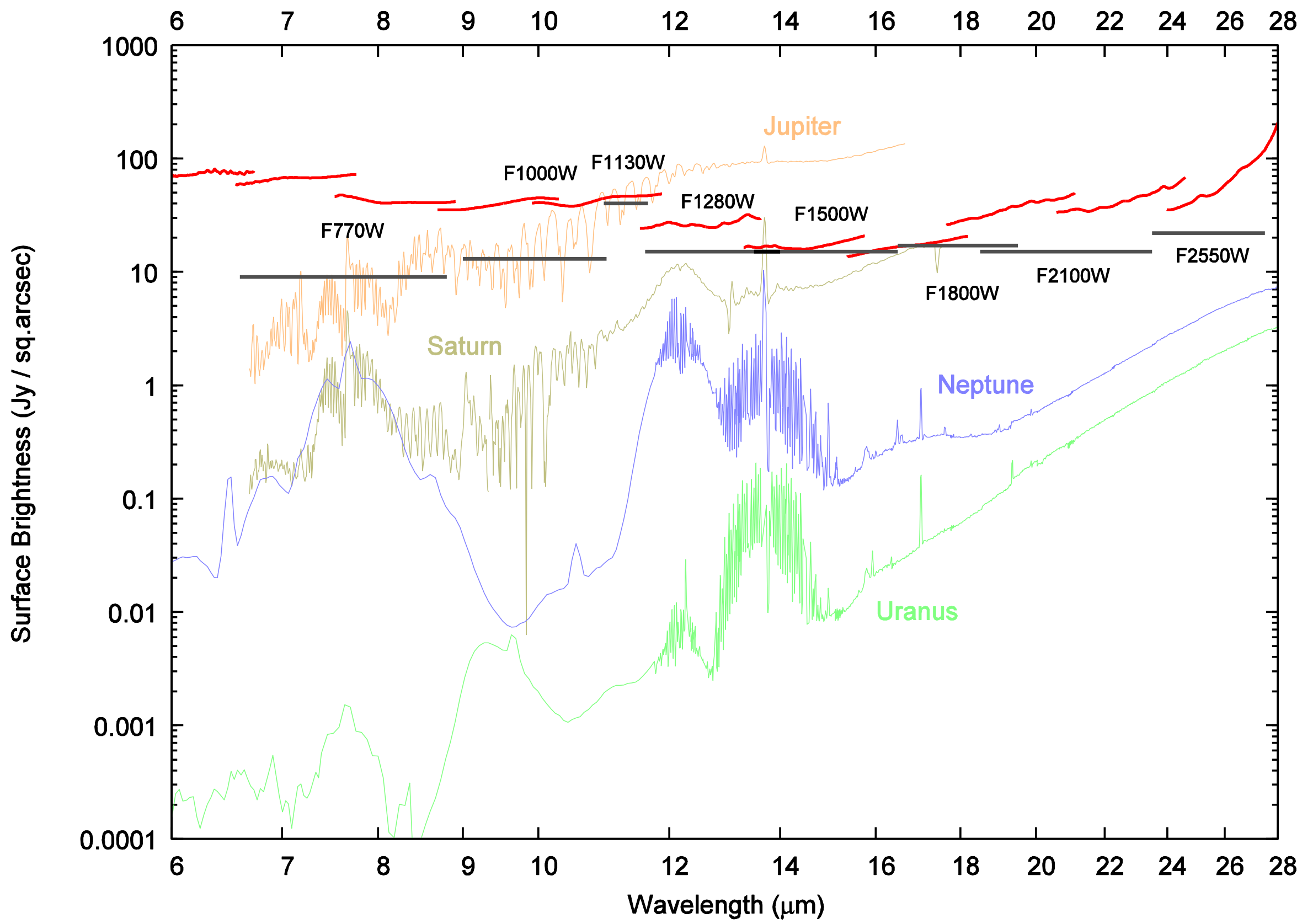

Figure 3. Mid-infrared spectra of the four giant planets compared to the saturation limits of filters on the MIRI imager (black lines with labels, F560W not shown) and Medium-Resolution Spectrometer (MRS, red curves). The imager saturation limits assume the smallest possible sub-array (SUB64). Jupiter will saturate in all mid-infrared imagery, while Saturn may be imaged in some MIRI filters; both planets will be observable with MRS at shorter mid-infrared wavelengths. Uranus and Neptune are observable via imagery and spectroscopy over MIRI's entire spectral range. Jupiter and Saturn spectra are from Cassini/CIRS (Fletcher et al. 2009) with spectral resolution *R*~400; Uranus and Neptune spectra are Spitzer/IRS data (*R*~600) from Orton et al. (2014). In both cases the displayed spectra are at resolutions substantially lower than that of MRS (*R*~2400).

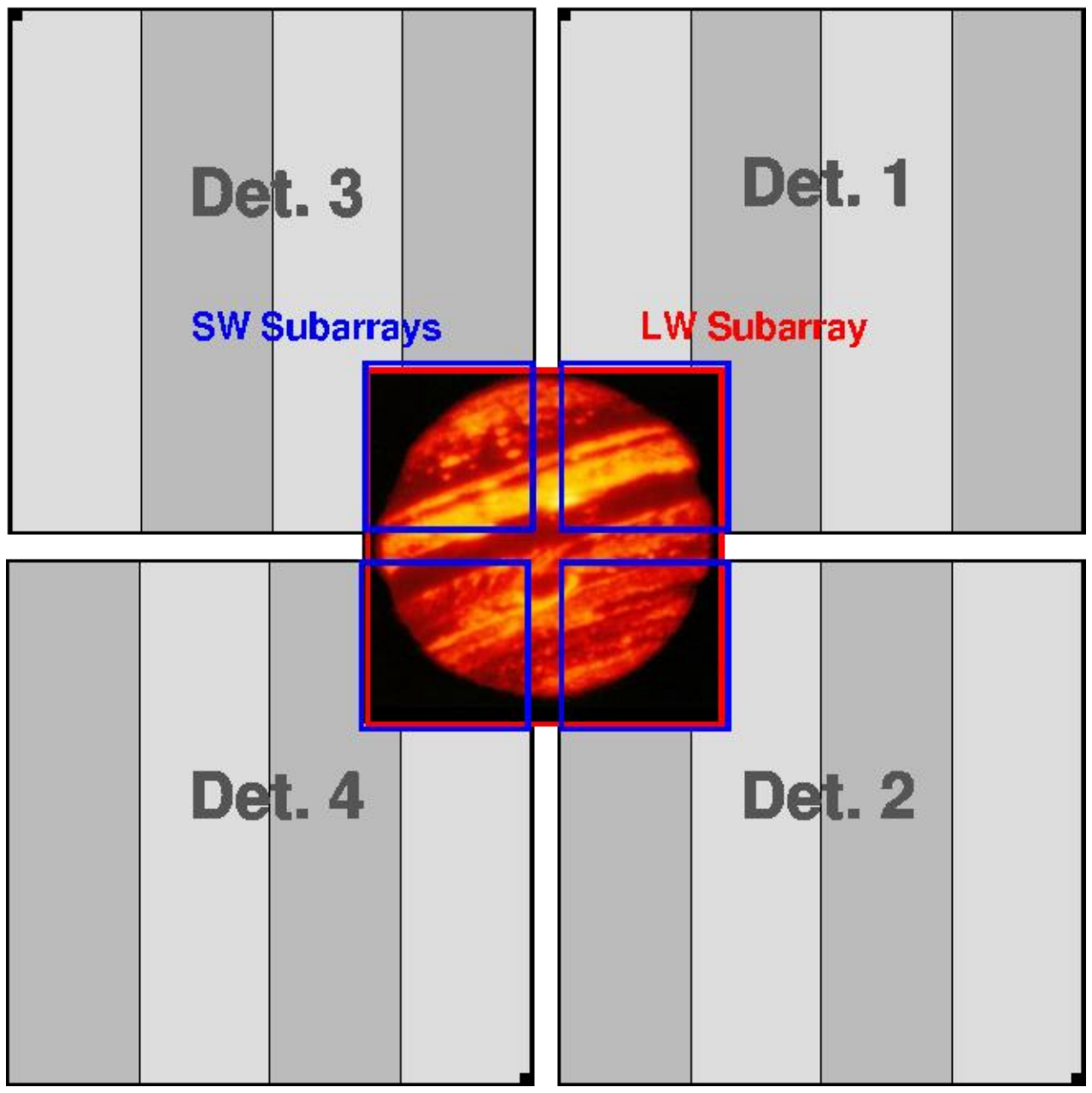

Figure 4. A possible configuration for observing Jupiter using sub-arrays on NIRCam. This sub-array (640×640 pixels) is of a size suitable for efficiently observing Jupiter, though smaller sub-arrays will be required for observing the planet in some NIRCam filters. Figure from Norwood et al. (2015).